\documentclass[
article,
superscriptaddress,
amsmath,
amssymb,
aps,
pra, 
longbibliography,
floatfix
]{revtex4-1}

\usepackage{setspace}

\usepackage{graphicx}
\usepackage{xcolor}
\usepackage{bm}
\usepackage[mathlines]{lineno}
\usepackage{tabularx}

\usepackage{float}

\usepackage{comment}
\usepackage{enumitem}

\usepackage{textcomp,mathcomp}

\usepackage{color}
\usepackage{dcolumn}

\usepackage[english]{babel}
\usepackage[latin1]{inputenc}
\usepackage[bottom=2.5cm,top=2.5cm, left=2cm, right=2cm]{geometry}
\usepackage{siunitx}
\usepackage{xr}
\makeatletter
\newcommand*{\addFileDependency}[1]{
  \typeout{(#1)}
  \@addtofilelist{#1}
  \IfFileExists{#1}{}{\typeout{No file #1.}}
}
\makeatother

\newcommand*{\myexternaldocument}[1]{
    \externaldocument{#1}
    \addFileDependency{#1.tex}
    \addFileDependency{#1.aux}
}

\myexternaldocument{Supplementary_Materials}

\usepackage[unicode=true, bookmarks=true, bookmarksnumbered=false, bookmarksopen=false, breaklinks=false, pdfborder={0 0 1}, backref=false, colorlinks=false]{hyperref}

\begin{document}

\setstretch{2.75}

\title{Polarizing ultrathin ferroelectric BaTiO$_3$ films through interfacial layer polarization}

\author{Ipek Efe}
\email[]{ipek.efe@mat.ethz.ch}
\affiliation{Department of Materials, ETH Zurich, CH-8093 Zurich, Switzerland}
\author{Edith Simmen}
\affiliation{Department of Materials, ETH Zurich, CH-8093 Zurich, Switzerland}
\author{Tobias Goldenberger}
\affiliation{Department of Materials, ETH Zurich, CH-8093 Zurich, Switzerland}
\author{Manfred Fiebig}
\affiliation{Department of Materials, ETH Zurich, CH-8093 Zurich, Switzerland}
\author{Nicola A. Spaldin}
\affiliation{Department of Materials, ETH Zurich, CH-8093 Zurich, Switzerland}
\author{Morgan Trassin}
\email[]{morgan.trassin@mat.ethz.ch}
\affiliation{Department of Materials, ETH Zurich, CH-8093 Zurich, Switzerland}

\date {\today}

\begin{abstract}

An important requirement for the integration of ferroelectric thin films into devices is deterministic control of the polarization state in films of only a few unit cells in thickness. Here, we utilize the charged atomic planes of (001)-oriented SmNiO$_3$ (SNO) buffer layers as a polarizing template to stabilize the polarization in ferroelectric BaTiO$_3$ (BTO) model system thin films. We show that an upwards (downwards) oriented polarization is achieved by selection of the [SmO]$^+$ ([NiO$_2$]$^-$) buffer termination. Most importantly, the charged atomic planes of SNO suppress the depolarizing-field-induced critical thickness in BTO, and we record the emergence of a net polarization in our BTO films from the first unit cell deposited. Our experiments, guided by density-functional-theory (DFT) calculations, further highlight the impact of charged defects on the polarizing effectiveness of the SNO buffer. Specifically, oxygen vacancies counteract the polarizing field of the negatively charged, [NiO$_2$]$^-$-terminated surface of the SNO buffer. Our findings provide important insights into the interplay of defect chemistry and polarizing interfaces to stabilize ferroelectric polarization down to the single-unit-cell limit.

\end{abstract}

\maketitle

\section{Introduction}

Ferroelectric materials, which exhibit spontaneous electrical polarization that can be controlled by an external electric field, are foreseen to support the development of next-generation energy-efficient nonvolatile memories \cite{muller_ferroelectric_2023} and computing schemes \cite{everschor-sitte_topological_2024}. Their device integration, however, requires overcoming the depolarizing-field-induced effects that can promote a critical thickness for the electrical polarization to emerge \cite{junquera_critical_2003,wang_ferroelectric_2010,stengel_origin_2006}. Over the past decades, efforts have focused on control of the polarization in thin films using epitaxial strain \cite{martin_thin-film_2016} and charge-screening environments \cite{zubko_interface_2011,junquera_topological_2023}. An emerging strategy to control the polarization in oxide-ferroelectric thin films takes advantage of the so-called ``layer polarization" \cite{spaldin_layer_2021,stengel_electrostatic_2011}, which arises from the alternating stacking of oppositely charged crystal planes in ionic crystals \cite{noguera_polar_2000,goniakowski_polarity_2007,stengel_electrostatic_2011,spaldin_layer_2021,tasker_stability_1979}. The lack of compensation for these charges at surfaces and epitaxial interfaces brings new degrees of freedom for the engineering of the electrostatic landscape of ferroelectric thin films \cite{spaldin_layer_2021}. For instance, tuning the layer polarization through atomic-termination engineering in oxide multilayers has enabled the deterministic control of the polarization direction in many ferroelectric materials, including BiFeO$_3$ \cite{yu_interface_2012,de_luca_nanoscale_2017,efe_happiness_2021,spaldin_layer_2021}, BaTiO$_3$ \cite{li_enhancement_2022}, PbTiO$_3$ \cite{strkalj_-situ_2020,gattinoni_interface_2020}, and PbZr$_x$Ti$_{1-x}$O$_3$ \cite{yu_interface_2012,sarott_controlling_2023, strkalj_-situ_2020}.
Beyond setting the polarization direction, first-principles calculations suggest that the layer polarization could even trigger a polarizing field and induce a spontaneous polarization in otherwise paraelectric materials, such as KTaO$_3$ \cite{gattinoni_prediction_2022}, or enforce a polar state in BaTiO$_3$ \cite{simmen_interplay_2025} down to the ultrathin regime. Such a polarizing field, therefore, presents a promising route to suppressing the detrimental consequences of the depolarizing field in thin films and stabilizing a strong polarization in ferroelectrics even in the ultrathin limit. However, a direct experimental observation of the layer-polarization-induced polarizing field and its robustness against the abundant charged defects commonly present in oxides has remained underexplored. 

Here, we combine in-situ optical second harmonic generation (ISHG) with ex-situ piezoresponse force microscopy (PFM), X-ray diffraction techniques, and first-principles DFT calculations to demonstrate the polarizing effect of the charged planes of SNO buffer layers on epitaxial BTO thin film model systems. We set an upwards or downwards oriented ferroelectric polarization in BTO by tuning the surface termination of the SNO layer to be either (Sm$^{3+}$O$^{2-}$)$^{+}$ ([SmO]$^+$) or (Ni$^{3+}$O$_2$$^{4-}$)$^{-}$ ([NiO$_2$]$^-$), respectively. Remarkably, through monitoring the polarization in-situ during growth, we find that the layer polarization of SNO at the BTO$|$SNO interface triggers an onset of polarization from the first unit cell of the deposited BTO film with upwards polarization. In contrast, we observe a critical thickness for polarization in downward-polarized BTO. Through oxygen-pressure-dependent in-situ polarization tracking experiments and DFT calculations, we attribute this striking difference to oxygen-vacancy defects suppressing the polarizing effect of the [NiO$_2$]$^-$-terminated SNO layer. Finally, by tuning the oxygen partial pressure during the deposition, we eliminate the critical thickness for both BTO polarization directions. Thus, with our work, we bring new insights into ferroelectric polarization control by crystal-chemistry engineering and provide a robust pathway to achieving ultrathin polarization down to the single-unit-cell limit.

\section{Results and Discussion}

We start our investigation by exploring the use of a buffer layer consisting of formally charged planes to control the polarization direction of a model ferroelectric system, BTO. Among the many possible systems \cite{efe_engineering_2024}, SNO stands out due to its large layer polarization. Along the [001]-growth direction, the crystal structure of SNO is characterized by the stacking of oppositely charged ionic planes of [SmO]$^+$ and [NiO$_2$]$^-$. In insulating SNO, this stacking leads to a formal layer polarization and corresponding surface charge density of $\sim$50 $\mu$C cm$^{-2}$, at the respective surfaces. A recent 
DFT study showed that, somewhat surprisingly, the effects of the layer polarization, which is formally defined only for insulating
systems, persist even in the presence of free carriers in the metallic phase \cite{simmen_interplay_2025}. Note that this layer polarization is larger than that of most commonly employed buffer materials with charged layers, such as LSMO ($\sim$35 $\mu$C cm$^{-2}$) \cite{yu_interface_2012,spaldin_layer_2021}, and also exceeds the intrinsic bulk BTO polarization ($\sim$26 $\mu$C cm$^{-2}$) \cite{jaffe_chapter_1971}. 

We investigate BTO$|$SNO bilayers grown epitaxially on (001)-oriented SrTiO$_3$ (STO) substrates. In the BTO$|$SNO bilayers, two distinct interfacial configurations can be engineered: positively charged [TiO$_2$$|$SmO]$^{+}$ and negatively charged [BaO$|$NiO$_2$]$^{-}$ interfaces (see Fig. 1a-b). 
The substrate-induced epitaxial compressive strain state of BTO ($\varepsilon$ $=$ $-$2.23\%) favors an out-of-plane polarization orientation with an enhanced value of $\sim$35 $\mu$C cm$^{-2}$ \cite{kim_critical_2005}. As a result, the electrostatic energy can be reduced when the BTO polarization points towards a negatively charged [BaO$|$NiO$_2$]$^{-}$ interface and away from a positively charged [TiO$_2$$|$SmO]$^{+}$ interface, as depicted in Fig.1a-b. 

%
%
%
%

We grow the BTO$|$SNO$||$STO heterostructures using pulsed laser deposition (see Methods). The growth rate and surface morphology are monitored in situ using reflection high-energy electron diffraction (RHEED), confirming a two-dimensional growth mode for all layers. We achieve the [BaO$|$NiO$_2$]$^{-}$ interface by using a commercially available surface-treated STO substrate with TiO$_2$ termination \cite{de_luca_nanoscale_2017,strkalj_-situ_2020}. To create the [TiO$_2$$|$SmO]$^{+}$ interface, we first deposit 3 unit cells of SrRuO$_3$ (SRO) on STO, prior to the SNO layer. The SRO self-terminates with the SrO plane \cite{rijnders_enhanced_2004} due to the high volatility of Ru$^{4+}$ ions and sets the stacking order so that the SNO deposition starts with the [NiO$_2$]$^-$ plane \cite{de_luca_nanoscale_2017,gradauskaite_control_2025}. The completion of the SNO unit cell, then, leads to a [SmO]$^{+}$ surface termination (see Methods). To avoid strain relaxation in BTO, we focus on the ultrathin thickness regime and terminate the film growth at a BTO thickness of $\sim$8 nm. We note that the $\sim$ 0.01 -- 0.03 $\Omega$$\cdot$cm resistivity of our SNO buffer layer, measured by 4-probe resistance measurement at room temperature, suffices to serve as a bottom contact for PFM characterization.
Post-deposition X-ray diffraction (XRD), shown in Fig. 1c, confirms the high crystalline quality and the expected (001)-oriented growth of our BTO films. The reciprocal space mapping around the STO 103 reflection ensures that for both interface configurations the BTO peak is positioned at an in-plane lattice constant equal to that of STO (with the same Q$_{||}$ value), indicating that both BTO films are fully strained to the substrate (Fig. 1d). The extracted BTO \textit{c}-lattice parameter is 4.09 {\AA}, and the \textit{c/a} tetragonality is 1.05. 

\begin{figure}[h]
\includegraphics[width=0.70\columnwidth, trim={0cm 0cm 0cm 0cm}]{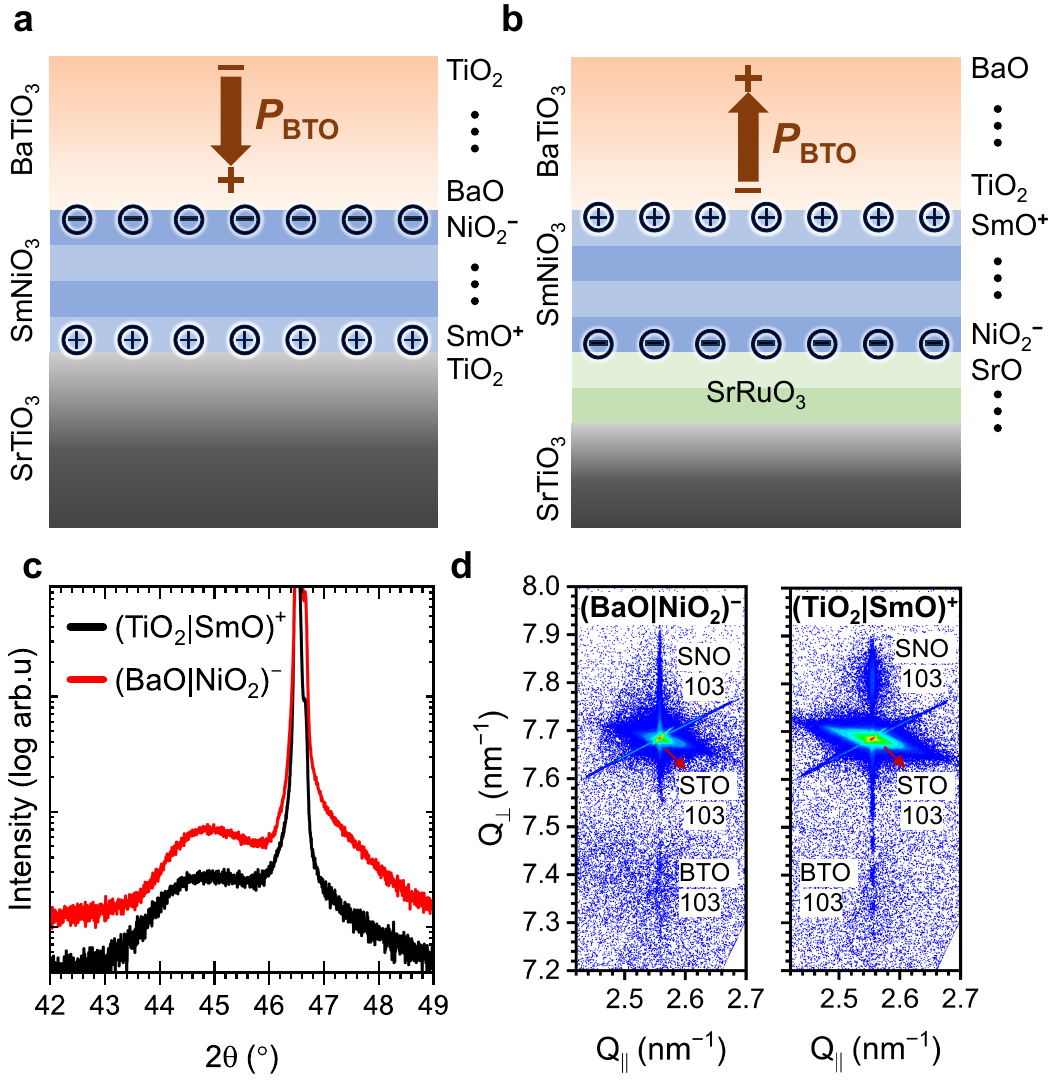}
\caption{\textbf{Interface termination engineering of BTO$|$SNO heterostructures} (a) BTO$|$SNO$||$STO with the [BaO$|$NiO$_2$]$^{-}$ interface and expected downward polarization in BTO. (b) BTO$|$SNO$|$SRO$||$STO with the [TiO$_2$$|$SmO]$^{+}$ interface and expected upward polarization in BTO. (c) XRD $\theta$/2$\theta$ scans of the BTO films with different interface configurations. A vertical offset was added to the XRD scans for clarity. (d) Reciprocal space mapping of the films showing fully strained BTO and SNO layers for both interface configurations. Note that the difference in the intensity of the SNO peaks stems from their different thicknesses. We have verified that SNO layers stay strained throughout the entire thickness range of 1.5 -- 13 nm that we have studied (see Supplementary Information, Fig. S1).}
\end{figure}

To investigate the impact of the charged SNO planes on the polarization states in our BTO films, we track the polarization during the BTO deposition using ISHG \cite{de_luca_nanoscale_2017,sarott_situ_2021}. In the electric-dipole approximation, the doubling of the frequency of light in a material can only occur when inversion symmetry is broken. This makes SHG an ideal tool for non-invasive investigation of ferroelectricity in materials \cite{nordlander_probing_2018,denev_probing_2011}. Note that the compressive epitaxial strain imposed by the STO substrate leads to an increase of the ferroelectric transition to well above the growth temperature (T$\mathrm{_C}$ $>>$ T$_\mathrm{growth} =700${\textdegree}C) \cite{strkalj_depolarizing-field_2019} and enables in-situ polarization monitoring during the deposition. We choose the polarization of the incident light to set our ISHG sensitivity to the out-of-plane component of the net polarization ({\textit{P}$\mathrm{_{net}}$}$\mathrm{^{OOP}}$), \cite{nordlander_probing_2018} and we calibrate the ISHG intensity to the film thickness by in-situ RHEED monitoring and post-growth X-ray reflectivity measurements.

The ISHG signal recorded during the growth of BTO(8 nm) on both [NiO$_2$]$^-$ and [SmO]$^+$ terminated SNO-buffered STO is shown in Fig. 2a and 2b, respectively. In both cases, we detect an onset of the ISHG signal during the deposition of BTO. The ISHG signal intensity increases continuously as the BTO deposition progresses, consistent with the behavior of a growing out-of-plane-polarized BTO film \cite{strkalj_depolarizing-field_2019}. Importantly, the SHG polarimetry performed after the growth, as shown in Figs. 2c and 2d, identifies an anisotropy which can be fitted by considering the non-zero SHG tensor elements of ferroelectric BTO with \textit{4mm} point-group symmetry (See Supplementary Information, Fig. S2). Post-deposition PFM experiments, see Figs. 2d and 2f, confirm the ferroelectric nature of the films and show that the orientation of the net out-of-plane polarization direction in our BTO films is determined by the termination of the SNO buffer layer.

\begin{figure}[h]
\includegraphics[width=0.65\columnwidth]{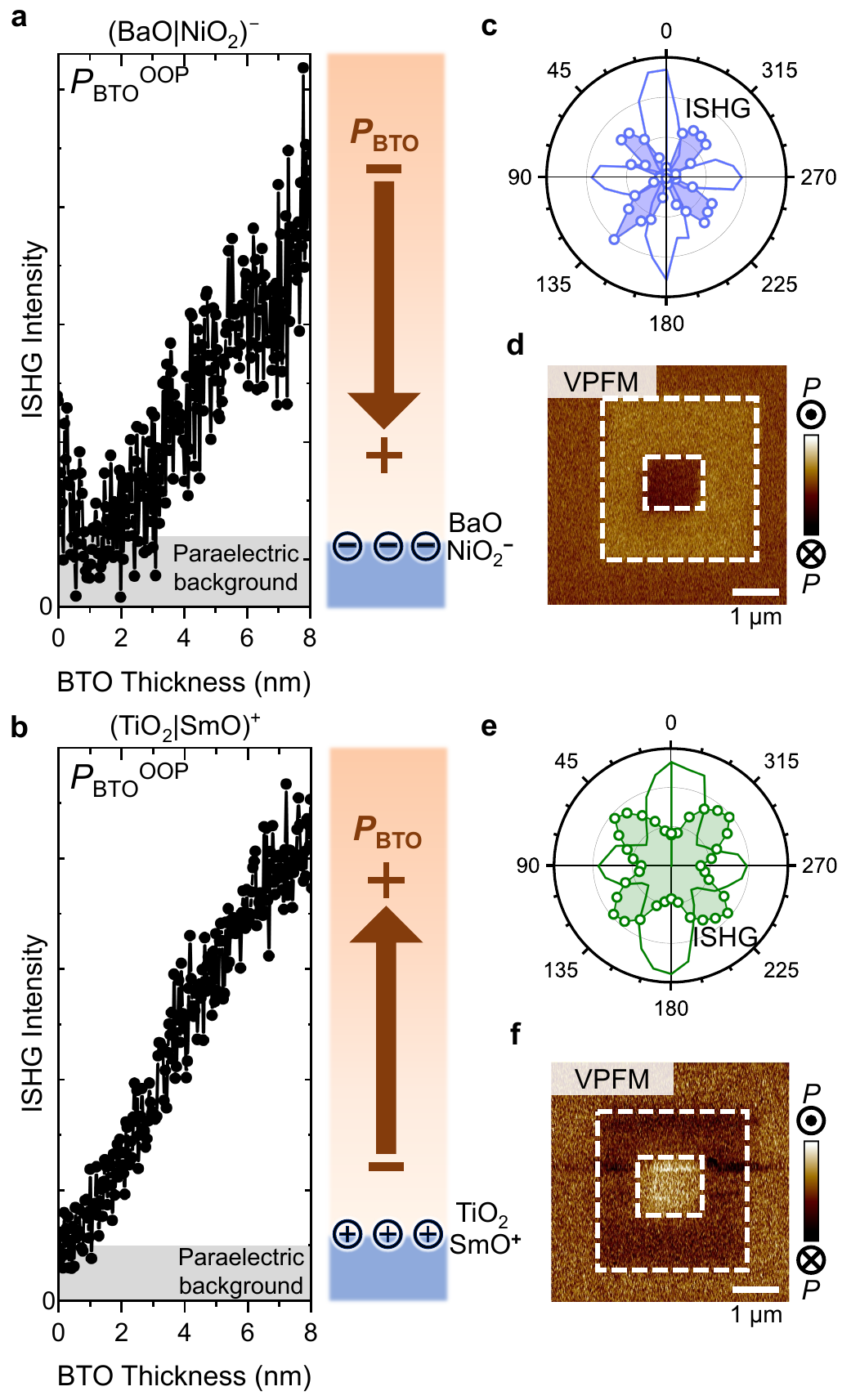}
\caption{\textbf{Interface dependence of critical thickness in BTO films.} (a-b) Evolution of the net out-of-plane polarization during the growth of an 8-nm BTO layer on (a) [NiO$_2$]$^-$-terminated SNO and (b) [SmO]$^{+}$-terminated SNO monitored by ISHG. The film in (a) indicates a critical thickness of $\sim$ 1.6 nm, while the film in (b) shows an onset of polarization as soon as the film growth starts. Note that the initial signal decrease in (a) originates from the contribution of the symmetry breaking at the newly created BTO$|$SNO interface, which interferes destructively \cite{strkalj_-situ_2020} with the previously created SNO$||$STO interface. When there is no contribution from BTO polarization to the signal, the destructive interference is observed as a decrease in the signal. On the other hand, in (b), due to the additional contribution from the BTO polarization dominating over the destructive interference of the interfaces, there is no initial decrease apparent. (c-e) The polarimetry scans of the heterostructure with the [NiO$_2$]$^-$-interface (c) and [SmO]$^{+}$-interface (e), acquired by rotating the polarizer while keeping the analyzer at \ang{0} (S-polarization, shaded plot) and \ang{90} (P-polarization, line plot). Polarimetry scans are consistent with the $4mm$ point-group symmetry (Supplementary Information). (d-f) The vertical-PFM (VPFM) contrast obtained upon the box-in-box poling of the BTO films with [BaO$|$NiO$_2$]$^{-}$ (d), and [TiO$_2$$|$SmO]$^{+}$ (f) interfaces.}
\end{figure}

Most strikingly, we notice a pronounced difference in the ISHG onset in the early stage of growth depending on the SNO termination. The BTO film grown on [NiO$_2$]$^-$-terminated SNO exhibits an onset of ISHG signal from $\sim$1.6-nm thickness. This value is consistent with the existence of a four-unit-cell critical thickness reported previously for BTO thin films grown on surfaces without layer charges \cite{strkalj_depolarizing-field_2019,tenne_ferroelectricity_2009}. Below this thickness, the depolarizing field, arising from the bound-charge accumulation induced by the BTO spontaneous polarization and pointing in the opposite direction, becomes dominant and suppresses the net polarization \cite{junquera_critical_2003}. When the [SmO]$^+$ termination of SNO is employed, in contrast, we observe an immediate onset of ISHG emission from the very first unit cell of BTO. This behavior was predicted by DFT calculations \cite{simmen_interplay_2025}, which showed that the layer polarization in SNO acts as an effective polarizing field and drives the out-of-plane polarization in the first unit cells of BTO. The difference in critical thickness between the two SNO terminations is surprising, however, since the magnitude of the layer polarization is the same in both cases \cite{simmen_interplay_2025}.

In order to shed light on the observed termination-dependent suppression of BTO critical thickness, we perform first-principles DFT calculations considering heterostructures consisting of 5 unit cells of BTO on SNO with a [TiO$_2$$|$SmO]$^{+}$ (Fig. 3a) and [BaO$|$NiO$_2$]$^{-}$ (Fig. 3b) interfacial chemistry.
The Ti displacements relative to the centrosymmetric position, shown in Fig. 3, indicate the local BTO dipoles pointing towards the surface (positive Ti displacement) or the BTO$|$SNO interface (negative Ti displacement). In both cases, the BTO layers exhibit a polarization. Moreover, the Ti displacements are larger in heterostructures with the  [BaO$|$NiO$_2$]$^{-}$ interface compared to the [TiO$_2$$|$SmO]$^{+}$ interface. This suggests that 
difference in interface chemistry alone cannot explain the termination-dependent polarization onset observed experimentally.

Next, we consider the interplay between the layer polarization and the defect chemistry, in particular, oxygen vacancies \cite{qiao_direct_2015,liu_local_2017,li_enhancement_2022,park_microscopic_1998}. Oxygen vacancies are commonly observed in SNO thin films, especially when they are tensile-strained \cite{zhang_unravelling_2022}. We introduce vacancies in DFT calculations by removing one neutral oxygen atom per heterostructure at a range of different lattice sites; in NiO$_2$ or SmO close to the interface, and in TiO$_2$ or BaO close to the interface or the surface. We compare the relative energy cost of introducing oxygen vacancies at different sites in BTO or SNO by calculating the difference in energy between the heterostructures with and without vacancy (\textit{E}$_{\mathrm{v_O^i}}$$-$\textit{E}$_\mathrm{no- {v_O}}$). We find that, regardless of the interface type, the oxygen vacancies lie preferentially within the NiO$_2$ layer closest to the interface (see Supporting Information, Fig. S3).

Strikingly, for the [TiO$_2$$|$SmO]$^{+}$ interface, oxygen vacancies enhance the Ti displacement inside BTO compared to the same heterostructure without the vacancy (Fig. 3a). In contrast, the vacancies at the [BaO$|$NiO$_2$]$^{-}$ suppress the Ti displacement, and hence the polarization inside the BTO (Fig. 3b). Since the absence of an O$^{2-}$ ion leaves a local positive charge \cite{das_synergistic_2019}, this positive charge adds to the layer charge density of the [TiO$_2$$|$SmO]$^{+}$ interface and enhances the polarizing effect of the SNO layer, while it reduces the net interfacial charge density of the [BaO$|$NiO$_2$]$^{-}$ interface and suppresses the polarizing effect of the SNO layer. Hence, the drastic difference in our BTO$|$SNO heterostructures with [BaO$|$NiO$_2$]$^{-}$ and [TiO$_2$$|$SmO]$^{+}$ interfaces is consistent with the calculated oxygen-vacancy accumulation.

\begin{figure}[h]
\includegraphics[width=0.85\columnwidth]{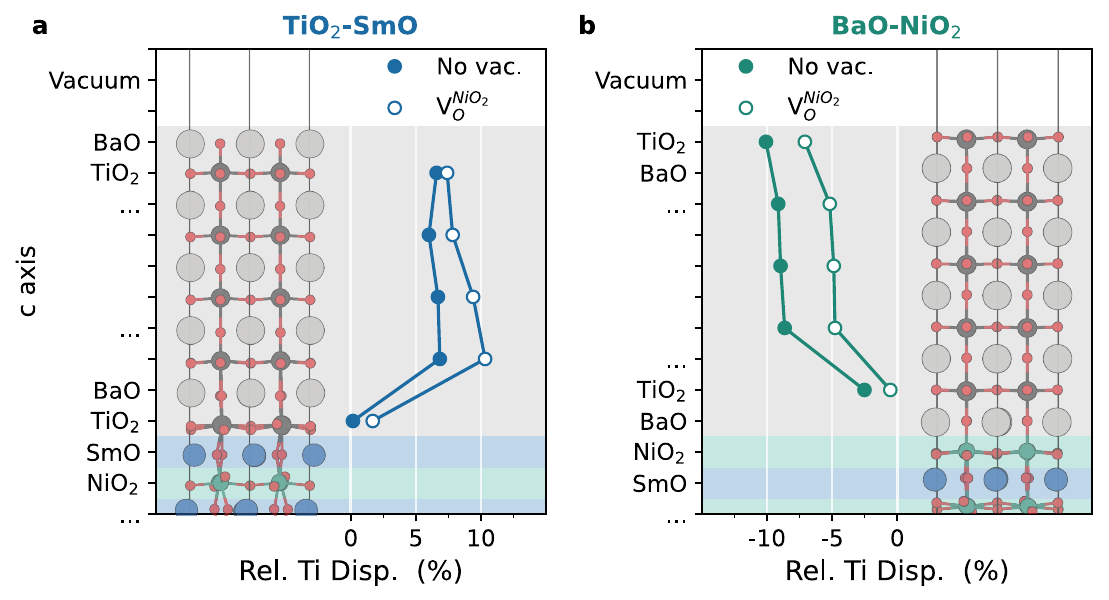}
\caption{\textbf{Interface-dependent suppression or enhancement of the BTO polarization by oxygen vacancies in DFT.} Ti displacements in the BTO slab relative to their centrosymmetric positions for the (a) [TiO$_2|$SmO]$^+$ and (b) [BaO$|$NiO$_2$]$^-$ interface for heterostructures with (empty circles) and without (full circles) an oxygen vacancy inside the NiO$_2$ sublayer closest to the interface. The relative Ti displacement is defined as $(z_{\text{Ti}} - c_\text{loc}/2) / c_\text{loc}$, where $z_\text{Ti}$ denotes the position of the Ti atom within the BTO layer and $c_\text{loc}$ is the height of the corresponding layer. The introduction of a vacancy has the opposite effect on the two interfaces: The polarization is enhanced for the [TiO$_2|$SmO]$^+$ interface while it is suppressed for the [BaO$|$NiO$_2$]$^-$ interface. The structure as relaxed in DFT without a vacancy is shown as an inset for both interfaces.
}
\end{figure}

To test our understanding of the interplay between oxygen vacancies and the polarizing effect of SNO on the ferroelectric polarization state of our film, we investigate the influence of the oxygen partial pressure during the BTO growth on our SNO-buffered STO substrates. Specifically, we increase the oxygen partial pressure tenfold, from 0.015 mbar to 0.15 mbar. This increase is expected to reduce the oxygen vacancy concentration in the BTO film and, hence, at the interface, restoring the polarizing effect of SNO and eliminating the critical thickness for BTO polarization for both interface configurations.
We again grow BTO films on two different SNO terminations and track the out-of-plane component of the polarization in situ by ISHG during BTO growth under 0.15 mbar oxygen partial pressure. 

The ISHG signals from BTO(7 nm)$|$SNO(8 nm)$|$SRO(1 nm)$||$STO and BTO(7 nm)$|$SNO(8 nm)$||$STO(TiO$_2$) are shown in Fig. 4a and 4b, respectively. The ISHG signal, whose anisotropy again could be fitted using the \textit{4mm} point-group-symmetry (see Supplementary Information, Fig. S4), is recorded for both films. In striking contrast with the films grown under lower oxygen partial pressure, this time, for both of the terminations, we detect an onset of ISHG, and thus of a net out-of-plane polarization, from the start of the BTO deposition. Note that the post-growth XRD analysis indicates a high crystalline quality despite the altered deposition conditions (Fig. 4c).

\begin{figure}[h]
\includegraphics[width=0.80\columnwidth]{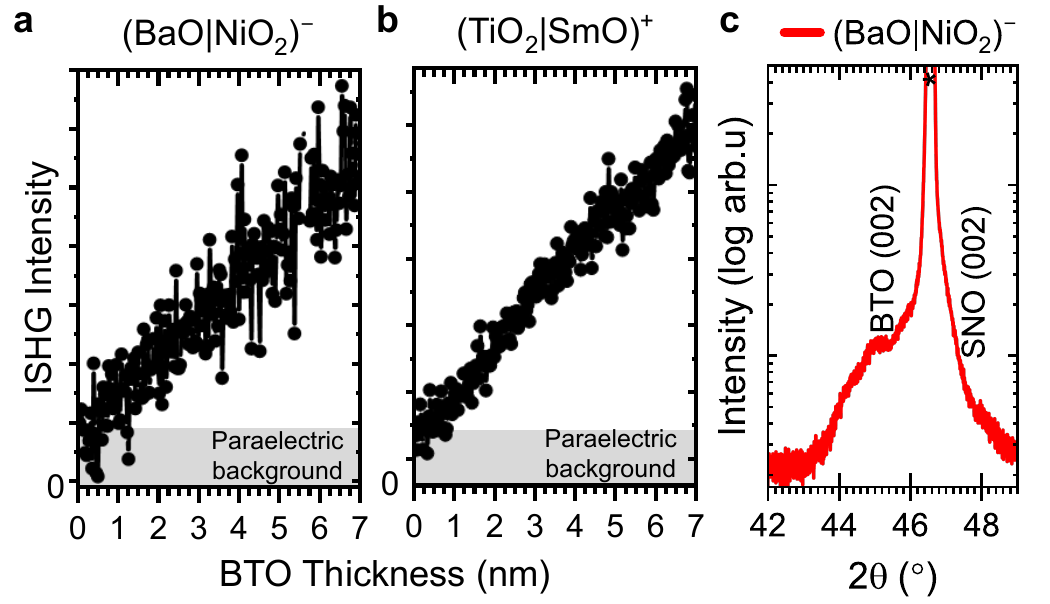}
\caption{\textbf{Eliminating critical thickness for both interfaces.} (a) Evolution of the net out-of-plane polarization during the growth of a 7-nm BTO layer on (a) [NiO$_2$]$^-$-terminated SNO layer and (b) [SmO]$^{+}$-terminated SNO layer monitored by ISHG. Both films show an increase in the ISHG signal directly with the start of the growth, indicating the absence of a critical thickness for both interfaces. (c) $\mathrm{\theta}$/2$\mathrm{\theta}$ scan of the BTO (7 nm)$|$SNO(8.5 nm)$||$STO(TiO$_2$) film, indicating the preservation of structural quality upon increasing the oxygen partial pressure during the deposition of BTO.
    \label{fig4.5:BTO_SNO_schematic}
}
\end{figure}

\section{Conclusion}

Our work demonstrates the use of SNO buffer layers to control the emergence of polarization in BTO layers. The layer polarization of SNO, which is created by its strongly charged [NiO$_2$]$^-$ and [SmO]$^{+}$ ionic planes, enables polarization direction control for the BTO and counteracts the detrimental impact of the depolarizing field. Specifically, it stabilizes the ferroelectric polarization in BTO thin films from the very first unit cell, providing a new avenue to combat depolarizing-field effects. In addition, we show that oxygen vacancies present in the system can have a strong influence on the effectiveness of the SNO polarizing field. Specifically, the oxygen vacancies suppress the polarizing field of the negatively charged [NiO$_2$]$^-$ interface while they enhance it for the positively charged [SmO]$^{+}$ interface. We find that increasing the oxygen partial pressure of the BTO deposition tenfold is sufficient to suppress the influence of oxygen vacancies and stabilize ferroelectricity in ultrathin BTO thin films for both polarization directions. Thus, here we provide a pathway to enforcing a polarization in the ultrathin regime, key for future integration of ferroelectric materials into nanoscale devices.
  
\section{Methods}
\subsection{Thin film growth and structural characterization}
The thin films and heterostructures were grown on STO (001) substrates (purchased from CrysTec GmbH) by pulsed laser deposition, using a KrF excimer laser at 248 nm. For the deposition of the films with (BaO$|$NiO$_{2}$)$^{-}$ interfaces, the TiO$_{2}$-terminated substrates were purchased as pre-treated. The thickness of the thin films was monitored using a combination of RHEED during growth and X-ray reflectivitiy ex-situ. 
XRD measurements to characterize the crystallinity, the thin-film orientation, and the epitaxial relationship between film and substrate were performed on a four-cycle thin-film diffractometer using Cu K$_{{\alpha}1}$ X-ray radiation (PanAnalytical X'Pert3 MRD and PanAnalytical X'Pert Pro Powder Diffractometer).
The SRO layers were grown at 700{\textdegree}C, with a laser fluence of 0.85 J cm$^{-2}$ and a repetition rate of 2 Hz. The O$\mathrm{_2}$ partial pressure was set to 0.015 mbar. The SNO layers were grown at 600{\textdegree}C, with a laser fluence of 0.8 J cm$^{-2}$ and a repetition rate of 4 Hz. The O$\mathrm{_2}$ partial pressure was set to 0.12 mbar. The BTO layers were grown at 650{\textdegree}C, with a laser fluence of 0.8 J cm$^{-2}$ and a repetition rate of 1 Hz. The O$\mathrm{_2}$ partial pressure was set to 0.015 mbar and 0.12 mbar for the films grown at low and high pressure, respectively.

\subsection{ISHG and SHG polarimetry measurements}

ISHG measurements were performed inside the pulsed-laser-deposition growth chamber in a reflection geometry under a 45{\textdegree} angle of incidence.  The probe beam of 1200 nm wavelength, with a pulse duration of 60 fs and a repetition rate of 1 kHz, was generated using an amplified Ti:Sapphire laser system. The pulse energy of the probe beam was set to 20 $\mathrm{\mu}$J. The beam was focused onto the sample choosing a probe diameter of 250 $\mathrm{\mu}$m. The SHG signal at 600 nm was selected using a monochromator and converted into an electrical signal with a photomultiplier tube and gate electronics \cite{de_luca_nanoscale_2017}. To capture the {\textit{P}$\mathrm{_{net}}$}$\mathrm{^{OOP}}$-related signal, the linear polarization of the incident probe beam and of the frequency-doubled SHG light were fixed to 90{\textdegree} (P-polarized) \cite{de_luca_nanoscale_2017}.

\subsection{Density functional calculations}
The DFT calculations were performed using the VASP code \cite{kresse_efficiency_1996, kresse_efficient_1996} with the exchange-correlation described by the Perdew-Burke-Ernzerhof functional revised for solids, PBEsol \cite{perdew_restoring_2008}. The pseudopotentials we used are based on the projector-augmented wave method \cite{blochl_projector_1994, kresse_ultrasoft_1999} with the following valence electron configuration: 5$s^2$ 5$p^6$ 5$d^{0.001}$ 6$s^{1.999}$ (Ba), 3$p^6$ 3$d^9$ 4$s^1$ (Ni), 3$s^2$ 3$p^6$ 3$d^3$ 4$s^1$ (Ti) and 2$s^2$ 2$p^4$ (O), 5$s^2$ 5$p^6$ 5$d^1$ 6$s^2$ (Sm), with the $f$ electrons frozen to the core for Sm. The heterostructures were constructed from $\sqrt{2} \times \sqrt{2}$ formula units of BTO and SNO in plane, and 4 formula units of SNO and 5 formula units of BTO out-of-plane. Oxygen vacancies were created by removing neutral oxygen atoms in the desired location. The NiO$_2$ sublayer adjacent to the interface was found to be the lowest-energy site for oxygen vacancies, see Supplementary Fig. S3. The in-plane lattice parameters were constrained to the lattice parameters $a = 3.898$ \AA of STO (calculated with DFT in Ref. \cite{wahl_srtio3_2008}) to simulate the substrate used in the experiments. To ensure that the interface and the surfaces do not interact, we included a 30 \AA vacuum region and used a dipole correction.  During the relaxation, the two SNO layers adjacent to the vacuum were frozen to their bulk structure and only the two layers next to the interface were allowed to relax. We used a cutoff energy of 650~eV and the internal coordinates were relaxed until the forces were smaller than $0.01~\text{eV\AA}^{-1}$. A $\Gamma$-centered zero-containing k-point mesh of ($8\times8\times1$) was used for the relaxations.

\subsection{Electrical transport measurements}
\subsection{Ex-situ atomic force microscopy and PFM measurements}
The surface topography, the ferroelectric polarization, and the local poling characteristics of the samples were studied using a Bruker Multimode 8 atomic force microscope equipped with Pt-coated Si tips (MikroMasch, \textit{k}= 5.4 N m$^{-1}$). The PFM measurements were performed using a drive voltage in the range of V$\mathrm{_{ac}}$ = 0.75 V -- 1.5 V at 10 kHz. The BTO films were poled by applying a bias voltage in the range of V$\mathrm{_{dc}}$= $\pm$ 5 -- 7 V to the scanning tip. 

\section{Acknowledgements}

M.T. acknowledges the Swiss National Science Foundation (SNSF) under project no.\ 200021\_236413. E.S. and N.A.S acknowledge funfing from ETH Zurich and the Swiss National Science Foundation (SNSF) under project no. 209454. The calculations for this work were performed on the ETH Zurich Euler cluster.

\section{Author Contributions}

I.E. performed the thin film growth, ISHG measurements, XRD characterization, scanning-probe measurements and analysis. E.S., T.G., and N.A.S. performed the DFT calculations. M.T. designed the experiment and supervised the work jointly with M.F. All authors discussed the results. The manuscript was written jointly by all authors.

\section{Data Availability}

The data that support the findings of this study are available upon reasonable request.

\newpage
\bibliography{REFs.bib}

\end{document}